\documentclass[aps,prl,twocolumn,superscriptaddress,reprint]{revtex4-1}
\usepackage[dvips]{graphicx}
\usepackage{dcolumn}
\usepackage{bm}
\usepackage{graphicx,color}
\usepackage[dvipsnames]{xcolor}
\usepackage{amsmath}
\usepackage{siunitx}
\usepackage{xcolor}
\usepackage{soul}

\newcommand{\edit}[1]{\textcolor{Black}{#1}}

\begin{document}

	\title{Efficient and Continuous Microwave Photoconversion in Hybrid Cavity-Semiconductor Nanowire Double Quantum Dot Diodes}

	\author{Waqar Khan}
	\affiliation{NanoLund and Solid State Physics, Lund University, Box 118, 22100 Lund, Sweden}
	\author{Patrick P. Potts}
	\affiliation{NanoLund and Mathematical Physics, Lund University, Box 118, 22100 Lund, Sweden}
	\author{Sebastian Lehmann}
	\affiliation{NanoLund and Solid State Physics, Lund University, Box 118, 22100 Lund, Sweden}
	\author{Claes Thelander}
	\affiliation{NanoLund and Solid State Physics, Lund University, Box 118, 22100 Lund, Sweden}
	\author{Kimberly A. Dick}
	\affiliation{NanoLund and Solid State Physics, Lund University, Box 118, 22100 Lund, Sweden}
	\affiliation{Center for Analysis and Synthesis, Lund University, S-221 00 Lund, Sweden}
	\author{Peter Samuelsson}
	\affiliation{NanoLund and Mathematical Physics, Lund University, Box 118, 22100 Lund, Sweden}
	\author{Ville F. Maisi}
	\email{ville.maisi@ftf.lth.se}
	\affiliation{NanoLund and Solid State Physics, Lund University, Box 118, 22100 Lund, Sweden}
		
	\date{\today}

\maketitle

\edit{\section{Abstract}}
Converting incoming photons to electrical current is the key operation principle of optical photodetectors and it enables a host of emerging quantum information technologies. The leading approach for continuous and efficient detection in the optical domain builds on semiconductor photodiodes. However, there is a paucity of efficient and continuous photon detectors in the microwave regime, because photon energies are four to five orders of magnitude lower therein and conventional photodiodes do not have that sensitivity. Here we tackle this gap and demonstrate how microwave photons can be efficiently and continuously converted to electrical current in a high-quality, semiconducting nanowire double quantum dot resonantly coupled to a cavity. In particular, in our photodiode device, an absorbed photon gives rise to a single electron tunnelling through the double dot, with a conversion efficiency reaching 6$\,\%$.

\edit{\section{Introduction}}

The last two decades have seen a\edit{n extensive} development of photodiodes in the optical regime~\edit{\cite{becker2005advanced,Reviewhadfield2009single,eisaman2011}}, driven by both a need to answer fundamental quantum optics questions but also to develop building blocks of key importance for emergent quantum technologies such as quantum key distribution~\cite{QKD1gisin2002quantum} and linear optics quantum computation~\cite{OQCo2007optical}. Major developments in the field that show promise to achieve the desired characteristics for such devices and uses, include the demonstration of near-unity photon-to-electron conversion efficiency~\cite{Juntunen2016} and high operation speed~\cite{PD2ackert2015high,Xia2009}. However, these developments have occurred in the optical domain and corresponding developments in the microwave regime are so far largely lacking. 

This is a major gap because the realization of an efficient, continuous microwave photodiode would eventually enable extending time-correlated photon counting to the microwave regime and electronics domain. It would also open for quantum technology applications for solid-state system read-out, such as photon correlation based measurements of qubit states~\cite{SPDonSCQopremcak2018measurement,Gu2017,Ghirri2020}. But realizing efficient and continuous photon detection in the microwave regime is challenging: while single-shot readout of single-electron tunneling events~\cite{gustavsson2007,Hanson2007,Pekola2013} as well as non-continuous pulsed photon detection via a superconducting qubit~\cite{Chen2011, Narla2016, Besse2018, Kono2018, SPDonSCQopremcak2018measurement, Gu2017} have been demonstrated, the continuous and efficient conversion of microwave photons to electric current has so far been missing.

In conventional optical photodiodes, a photon is absorbed by exciting an electron-hole pair over the semiconductor bandgap. The photodiodes’ pn-junctions then ‘split’ the electron-hole pairs to generate an electrical current that can be detected. However, in the microwave regime, this approach does not work because the photon energy is four to five orders of magnitude smaller than in the optical regime and suitable semiconductor materials with such small band gaps do not exist. Thus, to realize efficient and continuous photon detection in the microwave regime requires an alternative photodiode operation principle. 

Here we \edit{present} a photodiode device that is capable of efficiently and continuously convert microwave photons into electric current \edit{and measure the quantum efficiency of $\eta = 6\ \%$}. Together with the established single-shot detection of electrons~\cite{gustavsson2007}, our results pave the way for and continuous microwave detection with single-shot readout at the theoretically predicted unit quantum efficiency~\cite{vavilowong2017quantum}. \edit{The continuous nature of the demonstrated conversion process, similar to the optical photodiodes~\cite{becker2005advanced,Reviewhadfield2009single,eisaman2011}, opens up prospects to probe photons and their statistics beyond the gated-time regime~\cite{Chen2011, Narla2016, Besse2018, Kono2018, SPDonSCQopremcak2018measurement, Gu2017}.}

\section{Results}

\subsection{Device architecture and operation principle}
\begin{figure}[t]
	\begin{center}
		\centering \includegraphics[width=0.48\textwidth]{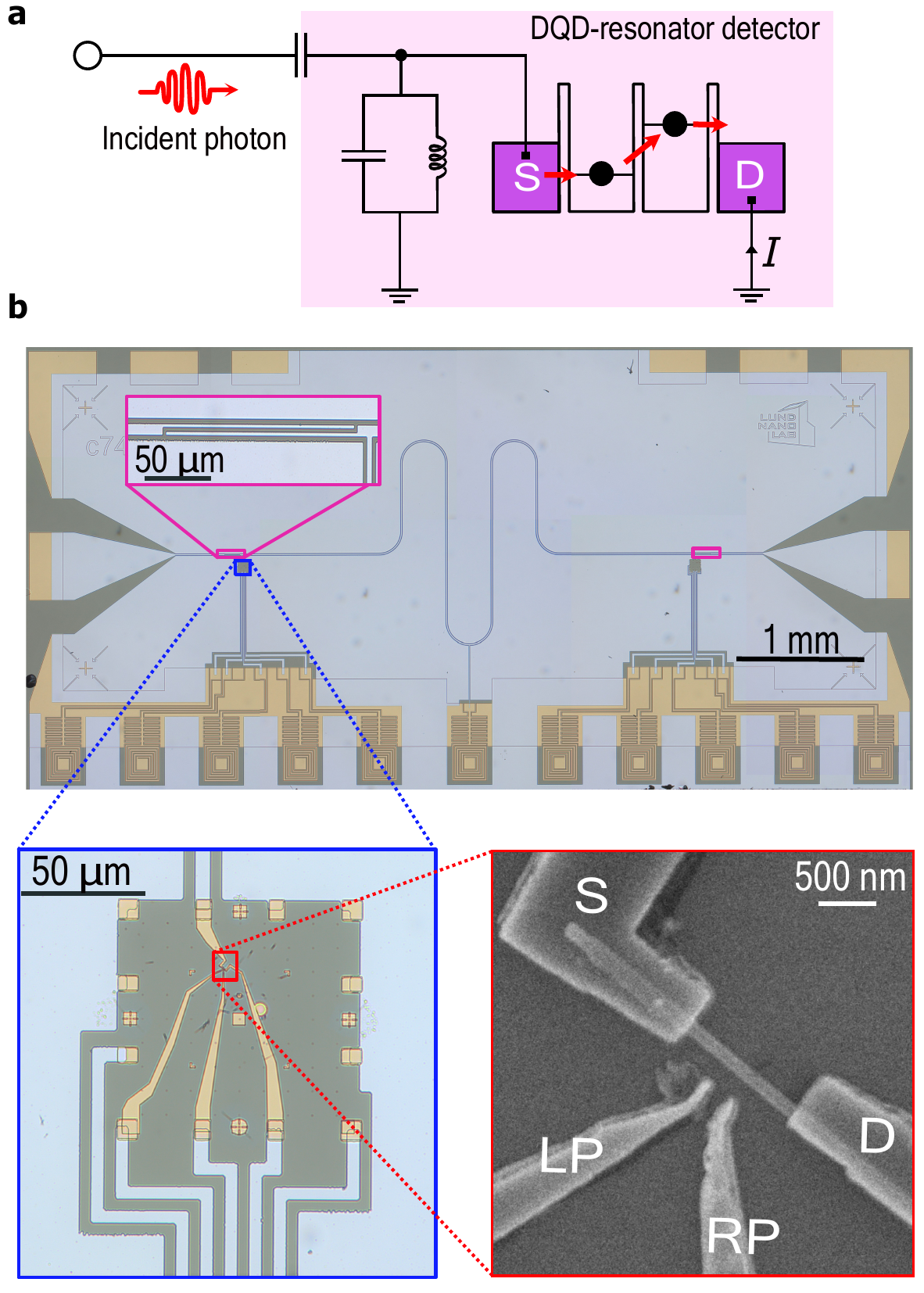}
		\caption{\textbf{Device architecture.} \textbf{a}, Schematic of \edit{the} detector operation: A photon in the transmission line is incident on the detector (pink shaded), consisting of a DQD embedded in a microwave resonator. With large efficiency, the photon is absorbed in the DQD, causing an electron to tunnel through the DQD, producing an electrical current $I$. {\bf b}, Optical image of the photodetector highlighting the superconducting microwave resonator made out of aluminum. The two zoom-ins (one optical and one scanning electron micrograph) depict the DQD and its connections to the resonator and measurement lines.}
		\label{deviceArchi}
	\end{center}
\end{figure}

Our photodiode device contains a crystal phase defined double quantum dot (DQD) in a semiconductor InAs nanowire~\cite{barker2019individually} embedded in a superconducting coplanar microwave resonator (Fig.~\ref{deviceArchi}). In our approach, the DQD forms an effective electronic two-level system with a tunable energy gap between the ground and excited states acting as an effective band-gap~\cite{stafford1996,oosterkamp1998,gustavsson2007,vanZanten2020}, which is analogous to optical pn-junction based detectors. During operation, an incident photon enters the cavity from a microwave port and is resonantly absorbed, exciting the DQD~\cite{vavilowong2017quantum}. As the ground (excited) state of the DQD is strongly localized on the left (right) dot, the photon absorption will result in an electron being transferred from source (left contact) to drain (right contact) with high probability. 

\subsection{Quantum efficiency}
\begin{figure}[t]
	\centering \includegraphics[width=0.4\textwidth]{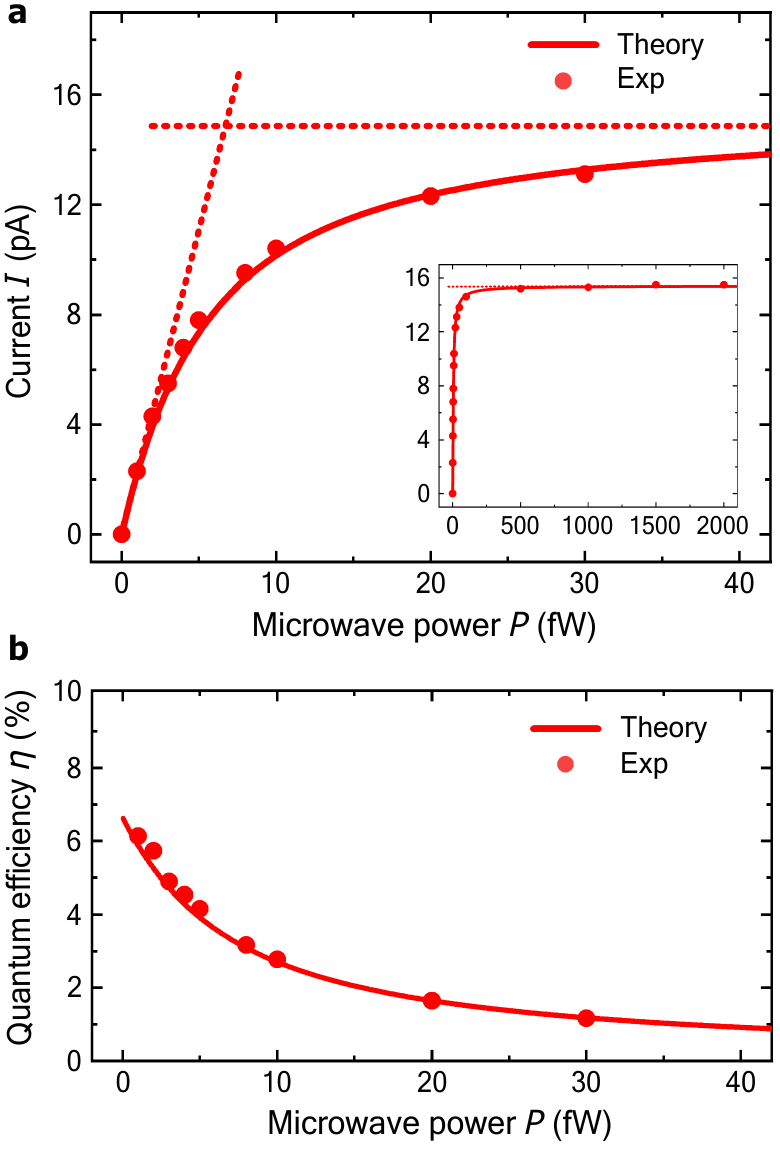}
	\caption{\textbf{Photoresponse and quantum efficiency.} {\bf a}, Photocurrent $I$ as a function of the ingoing microwave power $P$. Dots (solid line) present the measured data (theory) and the dashed lines are the theoretical predictions for the low power, linear response current (with $\eta = 6.6\,\%$) and the high power saturation current $I = 15\ \mathrm{pA}$. The incident microwave drive signal is in resonance with the resonator, $f=f_r$, and the DQD, $E=hf$, at a detuning $\delta = +\delta_r$ (c.f. Fig.~\ref{photodetection}). No bias voltage is applied. {\bf b}, Quantum efficiency, Eq. (\ref{eq:etaExp}), determined from the data of panel a. At $1\,$ fW input power, we experimentally reach $6\ \%$ efficiency. \edit{The abbreviation Exp denotes experimental data.}
	\label{QE}}
\end{figure}

\begin{figure*}[ht]
	\begin{center}
		\centering \includegraphics[width=.9\textwidth]{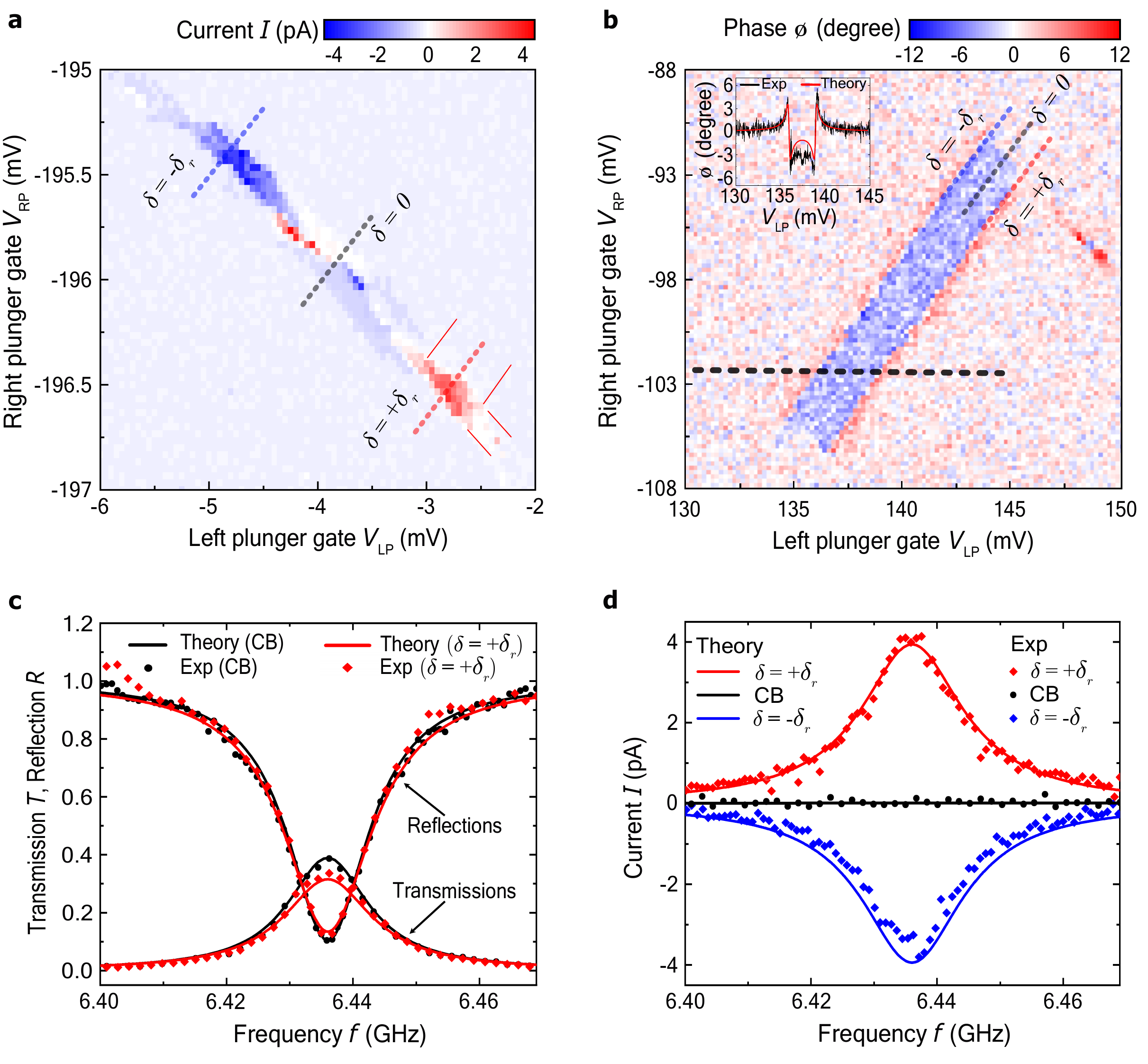}
		\caption{\textbf{Electronic and photonic response.} 
		\textbf{a}, Photocurrent $I$ as a function of the plunger gate voltages $V_{\rm{LP}}$ and $V_{\rm{RP}}$. The photocurrent shows peaks with opposite signs at detunings $\delta=\pm \delta_r$. The width of the peaks parallel and perpendicular to the lines of constant detuning (shown as dashed lines at $\delta=\pm \delta_r,0$) is given by the DQD energy gap $hf_r$ and $\tilde{\Gamma}+4g^2/\kappa=0.9\ {\rm GHz}$ respectively, as predicted by theory (thin solid lines at positive current peak). \textbf{b}, Phase shift $\phi$ of the resonator transmission as a function of $V_{\rm{LP}}$ and $V_{\rm{RP}}$, with constant detunings $\delta=\pm \delta_r,0$ shown as thin dashed lines. Inset shows cross-section at $V_{RP}= -102$ mV (along thick dashed line in main panel) together with theoretical fit. \textbf{c}, Transmission and reflection coefficients as a function of drive frequency $f$ at $\delta=\delta_r$ (red symbols, active photodetection) and for the DQD in the Coulomb Blockade (CB) regime, at $|\delta| \gg \delta_r$ (black symbols, no photodetection). \textbf{d}, Measured photocurrent as a function of $f$, together with theoretical fit, at $\delta=\pm \delta_r$ and at $|\delta| \gg \delta_r$ (CB regime). In all panels, the microwave power was $P = 2\ \mathrm{fW}$ and no source-drain voltage was applied. The measurements in panels a and b were made at resonance $f = f_r$. The plunger gate voltages in panels c and d are $V_{LP} = -2.8$ mV and $V_{RP} = -196.6$ mV for $\delta = \delta_r$ and $V_{LP} = -4.7$ mV and $V_{RP} = -195.4$ mV for $\delta = -\delta_r$. \edit{The abbreviation Exp denotes experimental data.}}		\label{photodetection}
	\end{center}
\end{figure*}

We realize a maximum efficiency $\eta = 6\,\%$ for our device\edit{. The high efficiency is } obtained thanks to the combination of the resonator enhancement of the microwave field in the vicinity of the DQD (which increases the photon-electron coupling~\cite{vavilowong2017quantum}) and the near-unity directivity for the photon-to-electron conversion in the DQD. We obtain this maximum efficiency for a microwave signal incident on the detector, with power $P=1$\,fW and frequency $f=6.436$ GHz, on resonance with both the fundamental resonator mode and the DQD energy gap. This induces a photocurrent $I= 2.4$ pA (see Fig.~\ref{QE}). The photon-to-electron conversion efficiency $\eta$ reads as
\begin{equation}
\label{eq:etaExp}
\eta \equiv \frac{hfI}{eP}=\frac{I/e}{\dot{N}}.
\end{equation}
Here $\dot{N}=P/hf$ is the rate of incident photons and $I/e$ the rate at which photoelectrons flow from source to drain. The experiment is modelled theoretically within a framework based on the Jaynes-Cummings Hamiltonian (see Methods). We find excellent agreement between experimental data and theory curves, see Figs. \ref{QE} and \ref{photodetection}.

\subsection{Energy detuning dependence}
The electronic and photonic response of the photodetector is shown in the four panels in Fig.~\ref{photodetection}. In Fig.~\ref{photodetection}a we present the photocurrent $I$ as a function of the plunger gate voltages $V_{\rm{LP}}$ and $V_{\rm{RP}}$, applied to the gates LP and RP (see Fig.~\ref{deviceArchi}) to move the energy levels of the quantum dots and thus change the level detuning $\delta$ and the energy gap $E = \sqrt{\delta^2+(2t)^2}$ of the DQD~\cite{Kowenandvan2002electron}. Here $t$ is the interdot tunnel coupling. For the DQD gap tuned in resonance with the resonator, $E=hf_r$ with $f_r=6.436$ GHz, the photocurrent displays peaks, as expected for photon assisted tunneling in DQDs ~\cite{stafford1996,oosterkamp1998}. This occurs at two detunings $\delta=\pm \delta_r$. The two current peaks have the same magnitude but opposite polarity, as anticipated from the symmetry of the DQD level configurations. Along the lines of constant detuning, $\delta=\pm \delta_r$, the photocurrent remains finite as long as the electrode chemical potential (same for source and drain) is between the ground and excited state of the DQD, giving a peak width $hf_r$. The extension of the photocurrent peaks perpendicular to the constant detuning line are due to resonance broadening and agree well with the theoretical low power prediction of $\tilde{\Gamma}+4g^2/\kappa$, where $\tilde{\Gamma}$ denotes the decoherence rate, $g$ the coupling strength of the DQD to the resonator, and $\kappa$ the resonator line-width.

Similarly to the photocurrent, the photonic response provides information on the DQD-resonator resonance conditions. The phase shift of the resonator transmission as a function of plunger gate voltages is shown in Figure~\ref{photodetection}b. The phase shift displays resonances symmetrically at detunings $\delta = \pm \delta_r$, with sharp transitions between positive and negative values, similar to previous DQD-resonator experiments ~\cite{petersson2012circuit, Liu2014,mi2017strong, highimpedenceReson_stockklauser2017strong, frey2012dipole, childress2004mesoscopic, delbecq2011coupling, deng2015coupling}. The phase shift resonances arise from interdot transitions without an electron tunneling to source or drain and are hence visible at $\delta=\pm \delta_r$, all along the interdot transition line.

\subsection{Frequency response}
The resonator reflection (R) and transmission (T) coefficients are shown in Fig.~\ref{photodetection}c as a function of drive frequency $f$ around $f_r$, for detunings $\delta=\delta_r$ and $\vert \delta \vert \gg \delta_r$. At $\vert \delta \vert \gg \delta_r$, when the DQD is in the Coulomb blockade (CB) regime and no photodetection takes place, the resonance lineshape is very well fitted with a Lorentzian with a central frequency $f_r = 6.436$ GHz and a linewidth of $15.5$ MHz [see also Eq.~\eqref{eq:trans_refl_cb} in the Methods section]. For $\delta=\delta_r$, at the positive photocurrent peak,  we observe the on\edit{-}resonance (i.e., when $f=f_r$) transmission decreasing by $\Delta T = -0.06 \, \pm\, 0.01$ and the reflection increasing by $\Delta R = 0.02 \, \pm\, 0.01$. As is clear from Fig.~\ref{photodetection}c, these changes are well described by theory. In fact, in the low power limit (see Methods), the active photodetector can be captured by an additional resonator loss term $\kappa_\mathrm{DQD}= 4g^2/\tilde{\Gamma}$, decreasing T and increasing R as photon absorption by the DQD opens up an additional channel for photons to escape the resonator.

We note that the additional amount of photons removed by the photodetector operation is given by the total change in transmission and reflection, i.e. the DQD decreases the outgoing photon flux by $\Delta T + \Delta R = -4\ \%$. Interestingly, the amount is comparable to but smaller than $\eta$. The difference arises from a fraction of the photons internally lost in the resonator, which are instead absorbed in the DQD when the detection is switched on, i.e. tuned from Coulomb blockade to the photodetection point: the photodetection reduces the fraction of photons internally dissipated in the resonator.

Figure~\ref{photodetection}d shows the photocurrent as a function of drive frequency $f$ for detunings $\delta=\pm \delta_r$ and $\vert \delta \vert \gg \delta_r$. At $\delta = \pm \delta_r$ the photocurrent lineshape, for both positive and negative current peaks, is within measurement accuracy the same as the one of the resonator response in Fig. 3c. We note that the detector bandwidth is given by the resonance linewidth. Indeed, in the low power-limit, theory predicts a Lorentzian with the same central frequency and width as for the transmission. This concurrence between the photocurrent and resonator response lineshapes provides direct evidence that the photodetector signal arises when photons entering the resonator are absorbed by the DQD, giving rise to an electrical current. At $\vert \delta \vert \gg \delta_r$, for the DQD in the CB regime, no photocurrent is observed as expected since the photodetector is tuned away from the operation point.

\subsection{Theoretical modeling}
The detector parameters are determined by fitting the measured results to the predictions of the full theory. As a first step, the bare resonator properties are determined with the DQD in the CB-regime. By fitting the full frequency dependence of the reflection and transmission coefficients, in Fig.~\ref{photodetection}c, the left and right resonator port couplings and the internal resonator losses are determined to $\kappa_{\rm L}/2\pi = 4.3\ \mathrm{MHz}$, $\kappa_{\rm R}/2\pi = 5.4\ \mathrm{MHz}$ and $\kappa_\mathrm{int}/2\pi = 5.8\ \mathrm{MHz}$ respectively. This gives the linewidth $\kappa=\kappa_{\rm R}+\kappa_{\rm L}+\kappa_{\rm int}=2\pi\cdot 15.5$ MHz.

Next, for the DQD-properties, we start by considering the directivity $p_f$, which is the fraction of photo-electrons that traverse the DQD from left to right, minus the fraction that traverse the DQD from right to left. For symmetric tunnel rates to source and drain, we find $p_f = \delta_r/hf_r$. We determine $\delta_r$ and the interdot tunnel coupling $t$ from the resonance conditions of Fig.~\ref{photodetection}b. The fit yields $t = 8.1\ \mathrm{\mu  eV}$ and $\delta_r = 21\ \mathrm{\mu eV}$, resulting in $p_f = 0.8$. Then, the tunnel rate $\Gamma /2\pi = 46\ \mathrm{MHz}$ is obtained from the high power saturated photocurrent $I = 2e \Gamma p_f/5$, see Fig. \ref{QE}. This tunnel rate also characterizes the dead time of the detector $\sim \Gamma^{-1} = \SI{20}{ns}$.

Finally, by fitting the data shown in Fig.~\ref{photodetection}b, c and d, we extract a DQD-resonator coupling of $g/2\pi = 21\ \mathrm{MHz}$ (corresponding to a bare coupling of $g_0/2\pi=38\ \mathrm{MHz}$, see Methods) influencing predominantly the strength of the phase response of Fig.~\ref{photodetection}b, the total decoherence rate of $\tilde{\Gamma}/2\pi= 790\ \mathrm{MHz}$, influencing its smearing, and the DQD relaxation rate of $\Gamma_r/2\pi = 23\ \mathrm{MHz}$ that, in addition to the other parameters, determines the low power photocurrent and the quantum efficiency of Fig.~\ref{QE}. With these parameters, we find excellent quantitative agreement between theory and measurements including both the resonator response as well as the photocurrent.
 
\edit{\section{Discussion}}
To analyze the possibilities for further improving the photodetector performance, we note from Figure \ref{QE} that the quantum efficiency is maximal in the low microwave power regime. Theory gives, in line with Ref.~\citealp{vavilowong2017quantum}, the low power efficiency
\begin{equation}
\label{eq:QElowP}
\eta = \frac{\kappa_{\rm L}}{\kappa}\frac{4\, \kappa_\mathrm{DQD}\, \kappa}{\left(\kappa_\mathrm{DQD}+\kappa\right)^2}\frac{\Gamma}{\Gamma+\Gamma_r}\, p_f.
\end{equation}
Hence, $\eta$ approaches unity when four conditions are met: First, the linewidth $\kappa$ needs to be dominated by the input port: $\kappa = \kappa_{\rm L}$. Second, the effective rate at which photons are absorbed by the DQD, $\kappa_\mathrm{DQD}$, should match the resonator linewidth: $\kappa = \kappa_\mathrm{DQD}$. Since $\kappa_\mathrm{DQD}=4g^2/\tilde{\Gamma}$, a larger coupling constant $g$ would allow unit efficiency with higher bandwidth (larger $\kappa$) and/or higher decoherence rate $\tilde{\Gamma}$. Third, the tunneling rate $\Gamma$ should be much larger than the DQD relaxation $\Gamma_r$. Fourth, the tunneling between the dots $t$ should be small compared to the photon energy $hf_r$ to obtain $p_f \rightarrow 1$. The second and the fourth condition have a trade-off because a small inter-dot tunnel coupling suppresses interaction with the resonator as $g\propto t$~\cite{childress2004mesoscopic}, and therefore reduces $\kappa_\mathrm{DQD}$. Unit efficiency is thus approached as the detector slows down. We estimate, from Eq.~(\ref{eq:QElowP}), that with an order of magnitude higher quality factor and source-drain tunnel coupling $\Gamma$, and a one-port cavity, $\kappa_{\rm L} = \kappa = \kappa_\mathrm{DQD} = 2\pi \cdot 2\ \mathrm{MHz}$, our photodetector would immediately reach a quantum efficiency of $\eta = p_f = 80\ \%$. Based on previous work on microwave resonators interacting with quantum dots, such high quality factor resonators and larger source-drain tunnel rates $\Gamma/2\pi \sim 1\ \mathrm{GHz}$ are experimentally obtainable~\cite{mi2017strong, petersson2012circuit, frey2012dipole} by adjusting the geometry of the input coupling capacitance and minimizing the internal losses of the resonator. This thus makes near-unity quantum efficiency detectors experimentally accessible.

Comparing our experiment to previous works, we \edit{demonstrated here} continuous photodetection with high efficiency in the microwave domain thanks to the combination of the resonator enhancement of the microwave field in the vicinity of the DQD, increasing the photon-electron coupling~\cite{vavilowong2017quantum}, and the high quality polytype DQDs with atomically sharp interfaces enabling a near-unity directivity for the photon-to-electron conversion. In addition, and of key importance for photocounting applications, the photodiode operation is continuous in time; a detected photon directly gives rise to an electron transfer through the DQD. This is in stark contrast to state-of-the-art single microwave photon detectors based on superconducting qubits~\cite{Chen2011, Narla2016, Besse2018, Kono2018, SPDonSCQopremcak2018measurement, Gu2017}, where photodetection, despite of reaching close to unity efficiency and demonstrating non-demolition capability, is indirect, occuring via qubit readout only at predetermined times.

Previous continuous detectors have predominantly focused on microwave spectroscopy and have an unknown, presumably orders of magnitude lower efficiency~\cite{stafford1996,oosterkamp1998,gustavsson2007,vanZanten2020}. In particular, the detection of single absorption events reported in Ref.~\citealp{gustavsson2007} obtains a photoelectron only for $\sim 10^{-7}\ \%$ of the generation events of the QPC emitter. This is seven orders of magnitude smaller than our $6\ \%$ efficiency. Furthermore, we unambiguously detect the photons that impinge on the detector while Ref.~\citealp{gustavsson2007} potentially features alternative channels such as phonons and plasmons. 

Moreover, the efficiency demonstrated here is two orders of magnitude higher than the one for photoemission in similar hybrid cavity structures~\cite{Liu2014,Stockklauser2015}. The low emission efficiency is explained by the dominance of DQD relaxation over photoemission, occuring at rates $\Gamma_r$ and $\kappa_{\rm DQD}$ respectively. For photodetection, the DQD relaxation instead competes with tunneling out of the DQD (rate $\Gamma$) which occur at comparable rates in our device. 

In conclusion, we have experimentally demonstrated conversion of microwave photons to electrical current and with that constructed a microwave photodetector. Combining our results with a standard single-shot charge readout together with high quality resonators opens up the avenue to build and perform microwave experiments with individual photons with high quantum efficiency.\\

\subsection{Data availability}
The data that support the findings of this study are available from the authors upon reasonable request. 

\edit{\subsection{Code availability}}
The numerical calculations were performed using QuTiP~\cite{qutip2012}.


\subsection{Acknowledgements}
We thank P. Krantz for useful discussions and acknowledge the financial support from NanoLund, Swedish National Science Foundation and the Knut and Alice Wallenberg Foundation through the Wallenberg Center for Quantum Technology (WACQT). P.P.P. acknowledges funding from the European Union’s Horizon 2020 research and innovation programme under the Marie Skłodowska-Curie Grant Agreement No. 796700.

\subsection{Author contributions}
V.F.M. conceived the experiment. S.L., C.T. and K.D.T. designed and fabricated the nanowires. V.F.M. and W.K. designed and fabricated the nanowire-cavity device. W.K. performed the measurements. W.K, V.F.M., P.P.P. \edit{and P.S.} performed the data analysis. P.P.P. performed the theoretical calculations and the numerical simulations. All authors contributed to the discussion of the results and the manuscript preparation. 

\edit{\subsection{Competing Interests}
The authors declare no competing interests.
}

\section{Methods}

\subsection{Device fabrication}
The microwave resonator is realized with superconducting Al/Ti layer on undoped Si/SiO$_2$ substrate with $\SI{200}{nm}$ thick SiO$_2$, using conventional photolithography techniques. The two ends of the resonator connect capacitively to microwave ports from which we send in photons as well as measure the outgoing photons. The DQDs were formed in InAs zinc-blende - wurzite heterostructures grown with metal organic vapor phase epitaxy. The
growth process is described in Ref.~\citealp{namazi2015selective} and the integration of the DQD to the device was done by following the recipe of Ref.~\citealp{barker2019individually}. Gold pads were used as interconnects to obtain conducting contacts between the resonator and the DQD lines. The coupling of the resonator and the DQD was achieved by connecting the voltage antinode of the resonator capacitively to the dipole moment of the inter-dot transition via the source electrode of the DQD~\cite{mi2017strong}. This connection allows for grounding the DQD at the the middle point of the resonator. The middle point of the resonator has a voltage node point and hence the connection does not distort the resonance~\cite{frey2012dipole,mi2017strong} while allowing for a finite photocurrent to flow through the structure without generating bias voltage and applying a bias voltage for the transport characterization. 

The device fabrication was made by starting with two photolithography steps were to fabricate metallic DC and interconnect pads of Ni/Au with thickness 5\:nm\:/\:95\:nm and microwave resonator from Ti/Al with thickness 5\:nm\:/\:200\:nm. After fabricating the resonator, the process continued as in Ref.~\citealp{barker2019individually}: InAs nanowires were transferred on the designated spots with pre-patterned alignement marks by pick and drop method. In the next step, electron-beam lithography was opted to connect one side of nanowire with the microwave resonator which serves as source resorvoir and the other end of nanowire is connected to a metallic DC pad, which acts as drain reservoir. In addition to source/drain contacts, gate electrodes are also formed using electron-beam lithography techniques. The electrode material for contacts is Ni with thickness $\SI{135}{nm}$.

\subsection{Measurements}
The measurements were preformed in a dilution refrigerator at a base temperature below 10 mK and electronic temperature of 40 mK. The measurements were started by characterizing the DQD with standard transport measurements (Supplementary Figure 1a) and the resonator by measuring transmission from left to right microwave port (Fig. 3d, CB). The resonator mode, i.e. frequency $f_r$ and linewidth $\kappa$, were determined from the transmission data. The tunnel couplings of the DQD were then tuned by choosing the electron numbers with the plunger gate voltages $V_{LP}$ and $V_{RP}$ such that the current in the cotunneling lines (marked gray in Supplementary Figure 1a) were of equal magnitude for the source and drain transition and the interdot coupling is small based on the phase response of Fig. 3b. These two conditions are obtained by tuning the difference and the mean of $V_{LP}$ and $V_{RP}$ respectively. After that, the measurements of the main article were performed with simple sweeps. Shifts in the voltages $V_{LP}$ and $V_{RP}$ arising from offset charge changes were compensated in the course of the measurement to keep the photodetector in the same operation point.

To calibrate ingoing and outgoing microwave power, we used the reflected power away from the resonance as a reference. In this case all the power is reflected away from the resonator allowing us to determine the cable losses and thus determine the ingoing and outgoing powers and transmission and reflection coefficients. \edit{The Supplementary Methods} presents the used microwave lines and lists the insertion losses and gains of the microwave components and describes the protocol and calculations of the calibration. 

We determine the transmission coefficient presented in the main article from the left port to right one, and the reflection coefficient by sending in a signal via microwave circulator to the right port and measuring it with the same amplification chain. Based on the reciprocity of the resonator response, the transmission from left to right port is identical to the reverse direction. Hence these measurement presents the reflection and transmission for a signal sent to the right port. Note that this conclusion holds even for a very asymmetric resonator. In addition, we have a rather symmetric system with $\kappa_{\rm L} \approx \kappa_{\rm R}$.

To determine the directivity $p_f$ of the photodetector we determined the detuning of the two resonance conditions $\delta = \pm \delta_r$ of Fig. 3b. We used the finite bias measurement of Supplementary Figure 1a as a reference to determine the lever arm for the detuning $\delta$ in response to the gate voltages. The calculations are presented in \edit{the Supplementary Methods}.

The linewidths of the data in Fig.~\ref{photodetection}c and d are: 
\begin{itemize}
    \setlength{\itemsep}{0pt}
    \setlength{\parskip}{0pt}
    \item Transmission in CB: $15 \pm 1$ MHz
    \item Reflection in CB: $17 \pm 2$ MHz
    \item Transmission, photodetection: $16 \pm 1$ MHz
    \item Reflection, photodetection: $17 \pm 3$ MHz
    \item Photocurrent, pos. polarity: $17 \pm 3$ MHz
    \item Photocurrent, neg. polarity: $19 \pm 3$ MHz
\end{itemize}
The linewidths and their uncertainty were determined by fitting a Lorentzian lineshape to the data. The fits yield the same linewidth within the fit uncertainty. All of the fits provide also the same resonance frequency of $6.436$ GHz within $1$ MHz uncertainty.

\subsection{Theory}
Theoretically, we describe the system using a Jaynes-Cumings Hamiltonian, given here in the frame rotating at the angular frequency of the incoming radition $\omega$ (here $\hbar=1)$
\begin{equation}
\label{eq:hamiltonian}
\hat{H}=\Delta_c\hat{a}^\dagger\hat{a}+\Delta_q\frac{\hat{\sigma}_z}{2}+ g\left(\hat{a}\hat{\sigma}^\dagger+\hat{a}^\dagger\hat{\sigma}\right)+\sqrt{\kappa_{\rm in}\dot{N}}\left(\hat{a}^\dagger+\hat{a}\right),
\end{equation}
where $\Delta_c=\omega_r-\omega$, with the angular frequency $\omega_r=2\pi f_r$, and $\Delta_q = \sqrt{\delta^2+4t^2}-\omega$, where $\delta$ denotes the difference in the on-site energies of the left and right dot. The Jaynes-Cummings coupling can be written in terms of the bare coupling as $g=g_02t/\omega_r$. The resonator is described by creation (annihilation) operators $\hat{a}^\dagger$ ($\hat{a}$), $\dot{N}$ denotes the rate of incident photons, and $\kappa_{\rm in}$ the loss rate of the input port. The DQD is described by a three level system provided by the bonding (ground) state $|g\rangle$, the anti-bonding (excited) state $|e\rangle$, each containing one extra electron in the DQD, and the empty state $|0\rangle$ where no extra electron resides in the DQD. Double occupancy of the DQD is prevented by Coulomb interactions. The spin matrices act on the subspace of ground and excited state, i.e., $\hat{\sigma}_z=|e\rangle\langle e|-|g\rangle\langle g|$ and $\hat{\sigma}=|g \rangle\langle e|$. The empty state has an energy equal to the chemical potential of the contacts, which is set to zero. We note that Eq.~\eqref{eq:hamiltonian} is only valid for detunings close the resonance value $\delta_r$ due to a rotating wave approximation.

Tunneling of electrons between the reservoirs and the DQD, as well as photon losses, relaxation, and decoherence are described by the Markovian master equation
\begin{equation}
\label{eq:master_equation}
\begin{aligned}
\partial_t\hat{\rho} =& -i[\hat{H},\hat{\rho}]+\Gamma\mathcal{D}[|0\rangle\langle e|]\hat{\rho}+2\Gamma\mathcal{D}[|g\rangle\langle 0|]\hat{\rho}\\&+\Gamma_r\mathcal{D}[\hat{\sigma}]+\Gamma_\phi\mathcal{D}[\hat{\sigma}_z]+\kappa\mathcal{D}[\hat{a}]\hat{\rho},
\end{aligned}
\end{equation}
with the superoperator $\mathcal{D}[\hat{A}]\hat{\rho}=\hat{A}\hat{\rho}\hat{A}^\dagger
-\frac{1}{2}\{\hat{A}^\dagger\hat{A},\hat{\rho}\}$. Here $\Gamma$ denotes the tunnel rate between source/drain and DQD (assumed equal for source and drain), $\Gamma_r$ is an internal relaxation rate (presumably induced by phonons), $\Gamma_\phi$ a decoherence rate (presumably induced by coupling to nearby charges) and $\kappa$ denotes the photon loss rate. Spin degeneracy is taken into account by the factor of two in the term that describes electrons entering the DQD. 

An electron which leaves the DQD upon absorbing a photon, i.e., a photo-electron, may both originate and end up either in the left or right contact. The corresponding tunnel rates are given by
\begin{equation}
\label{eq:rates}
\Gamma_{\rm R}^{\rm out} = \frac{\Gamma_{\rm L}^{\rm in}}{2} = \Gamma \frac{\omega_r+\delta_r}{2\omega_r},\hspace{.5cm}\Gamma_{\rm L}^{\rm out} = \frac{\Gamma_{\rm R}^{\rm in}}{2} = \Gamma \frac{\omega_r-\delta_r}{2\omega_r},
\end{equation}
where the terms $(\omega_r\pm\delta_r)/2\omega_r$ account for the localization of the wave functions, and the factor of two difference between in and out tunneling accounts for spin degeneracy. Note that since $\omega_r\gg k_B T$, electrons may only enter the DQD into the ground state and leave the DQD from the excited state. The charge current through the DQD may then be written as
\begin{equation}
\label{eq:current}
I=\langle e|\hat{\rho}|e\rangle e\Gamma_{\rm R}^{\rm out}-\langle 0|\hat{\rho}|0\rangle e\Gamma_{\rm R}^{\rm in},
\end{equation}
It is instructive to introduce the directivity, defined as the fraction of photo-electrons entering from the left and leaving to the right, minus the fraction entering from the right and leaving to the left
\begin{equation}
\label{eq:directivity}
p_f = \frac{\Gamma_{\rm L}^{\rm in}\Gamma_{\rm R}^{\rm out}}{2\Gamma^2} - \frac{\Gamma_{\rm R}^{\rm in}\Gamma_{\rm L}^{\rm out}}{2\Gamma^2}.
\end{equation}
For symmetric tunnel rates, as assumed here, the directivity reduces to the simple expression $p_f=\delta_r/\omega_r$.

In the low-drive limit $\dot{N}\rightarrow 0$, the density matrix may be expanded perturbatively to lowest order in $\sqrt{\dot{N}\kappa_{\rm in}}$ which results in the current
\begin{equation}
\label{eq:currlowp}
I = \frac{e\dot{N}16g^2\kappa_{\rm L}\tilde{\Gamma}p_f\Gamma/(\Gamma+\Gamma_r)}{16g^4+8g^2(\kappa\tilde{\Gamma}-4\Delta_c\Delta_q)+(\tilde{\Gamma}^2+4\Delta_q^2)(\kappa^2+4\Delta_c^2)},
\end{equation}
where we chose the left port as the input, $\kappa_{\rm in}=\kappa_{\rm L}$.
From this equation, the expression for the efficiency given in Eq.~\eqref{eq:QElowP} in the main text is recovered by setting $\Delta_c=\Delta_q=0$.
For $\Delta_c=0$, Eq.~\eqref{eq:currlowp} reduces to a Lorentzian (as a function of $\Delta_q$) with width $\tilde{\Gamma}+4g^2/\kappa$. This explains the width of the photocurrent peaks in Fig.~\ref{photodetection}a. For $\delta=\delta_r$ (i.e. $\Delta_c=\Delta_q=\omega_r-\omega$), and in the limit $\tilde{\Gamma} \gg \kappa+\kappa_{\rm DQD}$ [with $\kappa_{\rm DQD} = 4g^2/\tilde{\Gamma}$], Eq.~\eqref{eq:currlowp} reduces to a Lorentzian with width $\kappa+\kappa_{\rm DQD}$.

In the large drive regime, 
we can assume that the backaction of the DQD on the cavity, as well as fluctuations of the cavity field can be neglected and we replace
\begin{equation}
\hat{a}\rightarrow -\frac{2i\sqrt{\dot{N}\kappa_{\rm L}}}{\kappa}.
\end{equation}
This results in the current
\begin{equation}
\label{eq:currlargedrive}
I=e\frac{16g^2\kappa_{\rm L}\dot{N}\Gamma p_f}{\kappa^2\tilde{\Gamma}(\Gamma+\Gamma_r)+40g^2\kappa_{\rm L}\dot{N}}.
\end{equation}
Equation \eqref{eq:currlargedrive}
saturates at large drives to the value
\begin{equation}
\label{eq:satcurr}
I|_{\dot{N}\rightarrow\infty} = \edit{e}\frac{2\Gamma}{5}p_f.
\end{equation}

The theory curves shown in Figs.~\ref{QE} and \ref{photodetection}\edit{d} (at $\delta=\delta_r$) are obtained using Eq.~\eqref{eq:current} and solving the master equation numerically with $\kappa_{\rm in}=\kappa_{\rm L}$, except for the inset of Fig.~\ref{QE}a where Eq.~\eqref{eq:currlargedrive} was used to avoid dealing with large Hilbert space dimensions. The photocurrent at $\delta=-\delta_r$, is given by the negative of the current at $\delta=\delta_r$ due to symmetry.

To model the transmission and reflection coefficients, we use the input-output relations
\begin{equation}
\label{eq:inputoutput}
\langle\hat{b}_{\rm in,\alpha}\rangle+\langle\hat{b}_{\rm out,\alpha}\rangle = \sqrt{\kappa_{\rm \alpha}}\langle\hat{a}\rangle,
\end{equation}
with $\alpha = {\rm L, R}$. When the right port is used as the input, we obtain the reflection and transmission amplitudes
\begin{equation}
\label{eq:transnum}
\begin{aligned}
r &= \frac{\langle \hat{b}_{\rm out, R}\rangle}{\langle \hat{b}_{\rm in, R}\rangle} = i\sqrt{\frac{\kappa_{\rm R}}{\dot{N}}}\langle \hat{a}\rangle-1,\\
t&=\frac{\langle \hat{b}_{\rm out, L}\rangle}{\langle \hat{b}_{\rm in, R}\rangle} = i\sqrt{\frac{\kappa_{\rm L}}{\dot{N}}}\langle \hat{a}\rangle,
\end{aligned}
\end{equation}
where we used $\langle \hat{b}_{\rm in, R}\rangle=-i\sqrt{\dot{N}}$ and $\langle \hat{b}_{\rm in, L}\rangle=0$. The theory curves shown in Figs.~\ref{photodetection}b and c are obtained from Eqs.~\eqref{eq:transnum} and solving the master equation numerically with $\kappa_{\rm in}=\kappa_{\rm R}$.

In the low-drive limit at $\delta=\delta_r$ and under the assumption $\tilde{\Gamma}\gg\kappa+\kappa_{\rm DQD}$ we find
\begin{equation}
\label{eq:trans_refl}
\begin{aligned}
&T\equiv|t|^2 = \frac{\kappa_{\rm L}\kappa_{\rm R}}{\left(\frac{\kappa}{2}+\frac{\kappa_{\rm DQD}}{2}\right)^2+(\omega-\omega_r)^2},\\&R\equiv |r|^2 = \frac{\left(\frac{\kappa}{2}+\frac{\kappa_{\rm DQD}}{2}-\kappa_{\rm R}\right)^2+(\omega-\omega_r)^2}{\left(\frac{\kappa}{2}+\frac{\kappa_{\rm DQD}}{2}\right)^2+(\omega-\omega_r)^2}.
\end{aligned}
\end{equation}
We note that the transmission $|t|^2$ shows the same Lorentzian lineshape as the photocurrent in this limit.

In the Coulomb blockade regime (i.e., $|\delta|\gg\delta_r$), the DQD and the cavity are effectively decoupled and we model this regime by setting $g=0$ in Eq.~\eqref{eq:hamiltonian}. We find the standard expressions for the transmission and reflection ceofficients
\begin{equation}
\label{eq:trans_refl_cb}
\begin{aligned}
&T = \frac{\kappa_{\rm L}\kappa_{\rm R}}{\left(\frac{\kappa}{2}\right)^2+(\omega-\omega_r)^2},\\&R= \frac{\left(\frac{\kappa}{2}-\kappa_{\rm R}\right)^2+(\omega-\omega_r)^2}{\left(\frac{\kappa}{2}\right)^2+(\omega-\omega_r)^2}.
\end{aligned}
\end{equation}

\end{document}


\title{\edit{Supplementary Information -} Efficient and Continuous Microwave \edit{Photoconversion} in Hybrid Cavity-Semiconductor Nanowire Double Quantum Dot Diodes}

\author{W. Khan}
\affiliation{NanoLund and Solid State Physics, Lund University, Box 118, 22100 Lund, Sweden}
\author{P. P. Potts}
\affiliation{NanoLund and mathematical Physics, Lund University, Box 118, 22100 Lund, Sweden}
\author{S. Lehmann}
\affiliation{NanoLund and Solid State Physics, Lund University, Box 118, 22100 Lund, Sweden}
\author{C. Thelander}
\affiliation{NanoLund and Solid State Physics, Lund University, Box 118, 22100 Lund, Sweden}
\author{K. A. Dick}
\affiliation{NanoLund and Solid State Physics, Lund University, Box 118, 22100 Lund, Sweden}
\affiliation{Center for Analysis and Synthesis, Lund University, S-221 00 Lund, Sweden}
\author{P. Samuelsson}
\affiliation{NanoLund and mathematical Physics, Lund University, Box 118, 22100 Lund, Sweden}
\author{V. F. Maisi}
\email{ville.maisi@ftf.lth.se}
\affiliation{NanoLund and Solid State Physics, Lund University, Box 118, 22100 Lund, Sweden}

\date{\today}

\maketitle
\edit{\section*{Supplementary Methods}}

\subsection*{Determination of the directivity term and DQD energies}
The directivity of the DQD photodetector is given by the term $p_f = \cos\left(\theta\right) = \delta_r/hf_r = \sqrt{1 - (2t/hf_r)^2}$, where $\theta$ is the so-called mixing angle, $\delta_r$ the detuning at resonance, $t$ the interdot tunnel coupling and $f_r$ the resonance frequency of the resonator~\cite{vavilowong2017quantum,childress2004mesoscopic}. To determine the directivity, we use the value $f_r = \SI{6.436}{GHz}$ obtained from the resonator response. To obtain the other parameter value, the detuning $\delta_r$ in resonance, we first determine how the left gate electrode changes the detuning $\delta$ based on finite bias transport measurements. Then based on identifying the detuning when the DQD is in resonance with resonator yields us both the detuning and the interdot coupling $t$ in the photodetector operation point. We present below the measurements and the procedure of these experiments in detail.

\begin{figure*}[ht]
	\begin{center}
		\centering \includegraphics[width=0.95\textwidth]{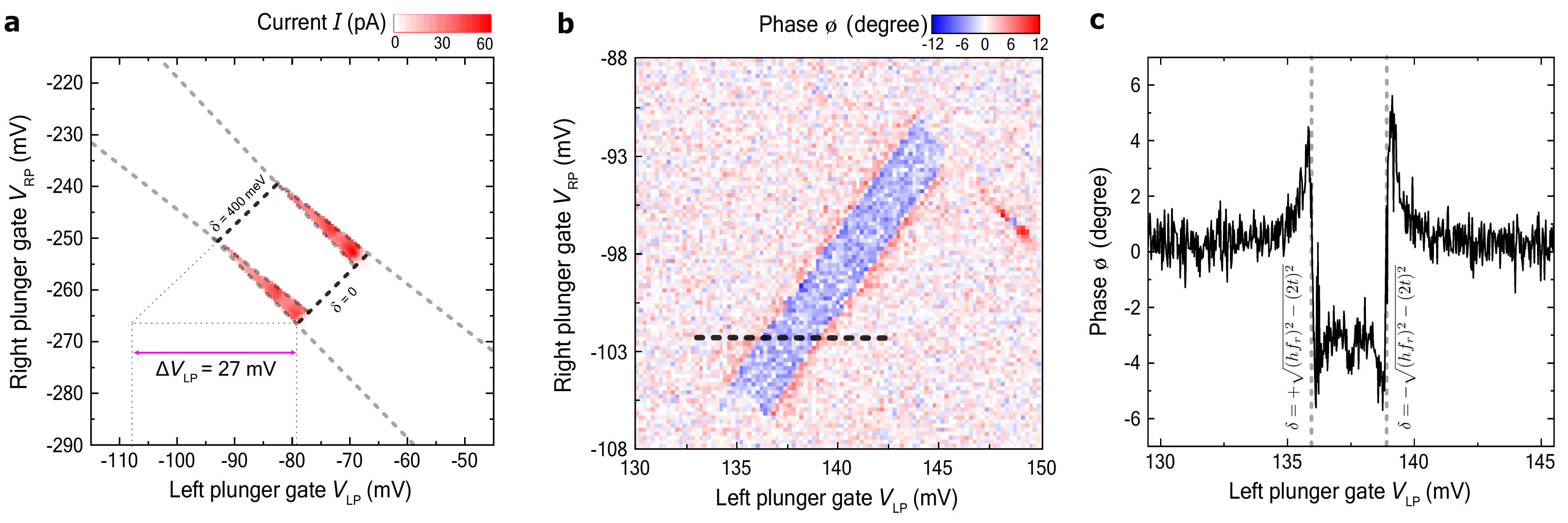}
		\caption{\textbf{a}, A pair of finite bias triangles from the charge stability diagram of DQD at $V_B = \SI{400}{\micro V}$. \textbf{b}, Phase response at interdot charge transition regime as a function of plunger gate voltages $V_{LP}$ and $V_{RP}$. \textbf{c}, Phase response cut through the interdot charge transition (dashed line in panel b). The DQD had a few charge reconfigurations between the measurements (sudden jumps in the gate voltages). By measuring the finite bias triangles (fingerprint from excited states and the electrical current $I$ profile), we assured that we were investigating the same transition at all times.}
		\label{tc}
	\end{center}
\end{figure*}

In Supplementary Fig.~\ref{tc}a, we present a finite bias measurement of the DQD. The bias voltage is set to $V_B = \SI{400}{\micro V}$ via the resonator middle contact. From the finite bias triangles, we identify two detuning conditions~\cite{van2002electronTransport}: at the base of the triangles the detuning vanishes, that is $\delta = 0$ and the peaks of the two neighboring triangles define a detuning line corresponding to bias voltage, i.e. $\delta = eV_B = \SI{400}{\micro eV}$. These values are the minimum and maximum detuning values such that the energy levels are within the transport window and dc current flows. Since the gate voltages move the energy levels of the quantum dots up and down with a linear relation, measuring the distance in the left gate voltage direction denoted as $\Delta V_{LP}$ in the figure, yields us the lever arm $\alpha_{{LP},\delta} = eV_B/\Delta V_{LP} = \SI{400}{\micro eV}/\SI{27}{mV}= \SI{14.8}{\micro eV/mV}$. This lever arm characterizes how much the left gate voltage shifts the detuning $\delta$ of the DQD. Here we have neglected the interdot tunnel coupling $t$ contribution to the total energy $E\left(\delta \right) = \sqrt{\delta^2+(2t)^2} \approx \left| \delta \right|$. This approximation is well justified since $e V_B \gg t$ as we will see next.

Now we determine the $\delta_r$ and the interdot tunnel coupling based on the resonance condition $E(\delta_r) = hf_r$. Here we use the full expression $E\left(\delta \right) = \sqrt{\delta^2+(2t)^2}$ without the above approximation since with the resonator phase response to the DQD we are able to measure the small deviance that $t$ causes to the approximate relation used elsewhere in our data analysis. Supplementary Figures~\ref{tc}b and c, present the phase response of the resonator as a function of the DQD gates. We observe a response along the direction where the detuning changes while the response stays constant along the perpendicular direction similarly as in previous experiments~\cite{frey2012dipole,vacuurabitoida2013vacuum,mi2017strong}. We observe the resonance of the DQD with the microwave resonator as a sharp transition in the phase response. We have one resonance for positive detuning and one for negative detuning. Panel c shows a cut of the response with the left gate voltage $V_{LP}$. With that, we measure the distance of the positive and negative resonance condition $\Delta V_{LP} =  \SI{2.84}{mV}$. The resonance conditions $E(\pm \delta_r) = hf_r$ result in a sharp transitions from a positive phase to a negative in the phase response. With the above lever arm, we convert this to the detuning at the resonance $\delta_r = \alpha_{{LP},\delta} \: \Delta V_{LP} /2 = \SI{14.8}{\micro eV/mV} \cdot \SI{2.84}{mV}/2 = \SI{21.0}{\micro eV}$ yielding us the directivity $p_f = 0.79$. With the total energy difference of the DQD in the resonance, $E\left(\delta_r \right) = \sqrt{\delta_r^2+(2t)^2} = hf_r$, we also obtain the tunnel coupling as $t = \sqrt{(hf_r)^2-\delta_r^2}/2 =  \SI{8.1}{\micro eV}$. 

In Supplementary Fig.~\ref{tc}a, we see the first excited state with pronounced current in the middle of the triangles. The bias voltage opens up an energy window of $\SI{400}{\micro eV}$. As the excited state is in the middle of it, we estimate the higher excited states of the quantum dots to be approximately $\SI{200}{\micro eV}$ above the states considered for the photodetector. Since this energy is an order of magnitude larger than the photon energy, and other energies in the system such as the thermal energy $k_BT$, the excited states of the quantum dots do not influence the photodetector operation.

Similar to the above determination of the detuning energies, we also determine the energy shifts along the detuning axis. This direction is identified by the base of the finite bias triangle where the energy difference $E$ of the ground and excited state is constant and the states are move up or down in energy in concert. The length of the triangle base in Supplementary Fig.~\ref{tc}a at $\delta = 0$ corresponds to an energy shift of $\SI{400}{\micro eV}$ of the DQD energy levels. This allows us to determine the source-drain energy window of $E = hf_r$ presented in Fig. 3a of the main manuscript.

\begin{figure}[t]
	\begin{center}
		\centering \includegraphics[width=0.35\textwidth]{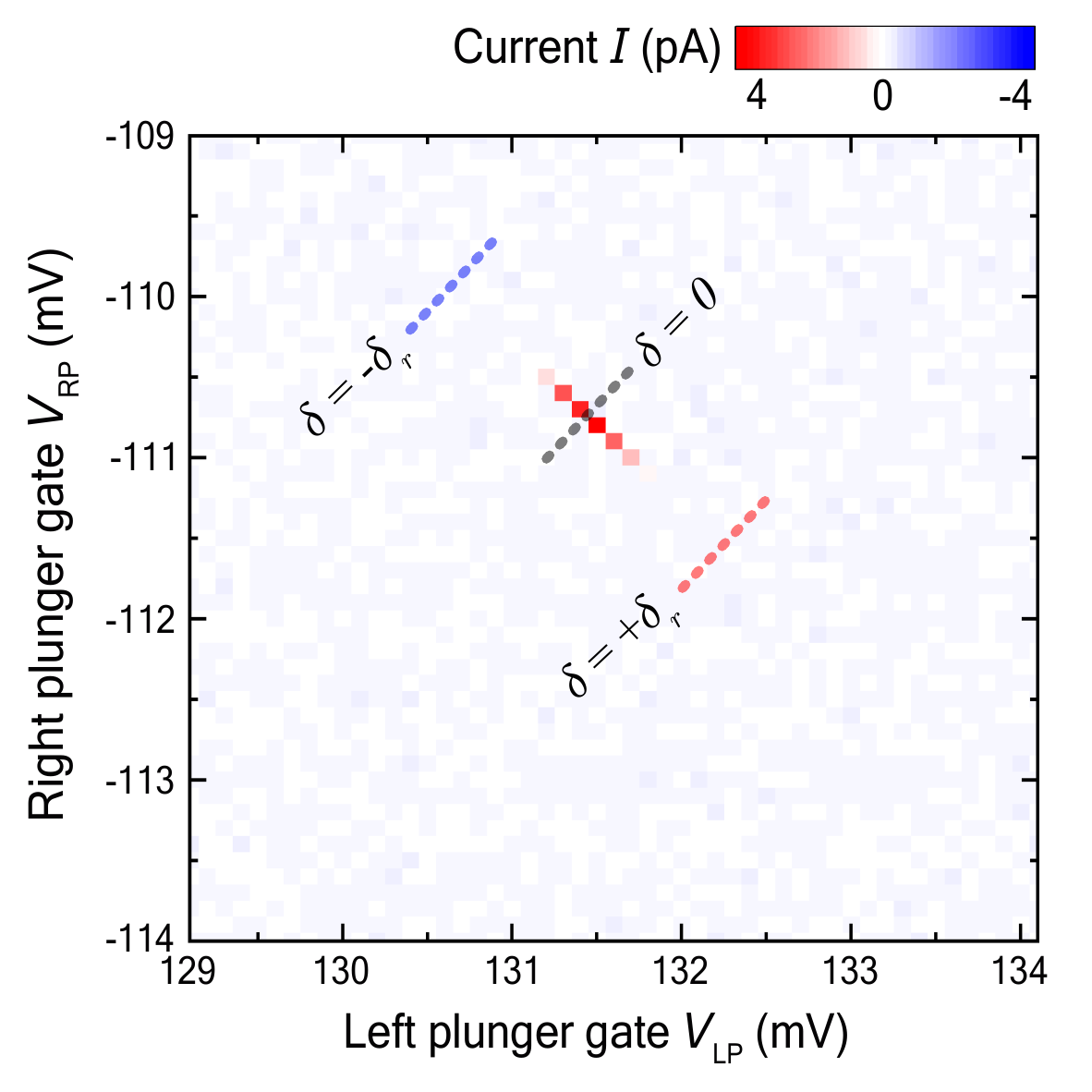}
		\caption {Photodetector response without applied microwave signal. No measurable signal is observed at the photodetector points. Electrical current flows only at a minuscule area around the triple point with $\delta = 0$.}
		\label{zeropower}
	\end{center}
\end{figure}

\edit{\subsection*{Dark current}}
Supplementary Figure~\ref{zeropower} presents the photodetector response with no applied microwave signal. We observe a miniscule signal at the charge triple point at $\delta = 0$. At the photodetector points $\delta = \pm \delta_r$, we see no measurable current. Thus, the dark current of the photodetector is below the 0.2 pA noise level of the measurement.

\begin{figure}[t]
	\begin{center}
		\centering \includegraphics[width=0.99\textwidth]{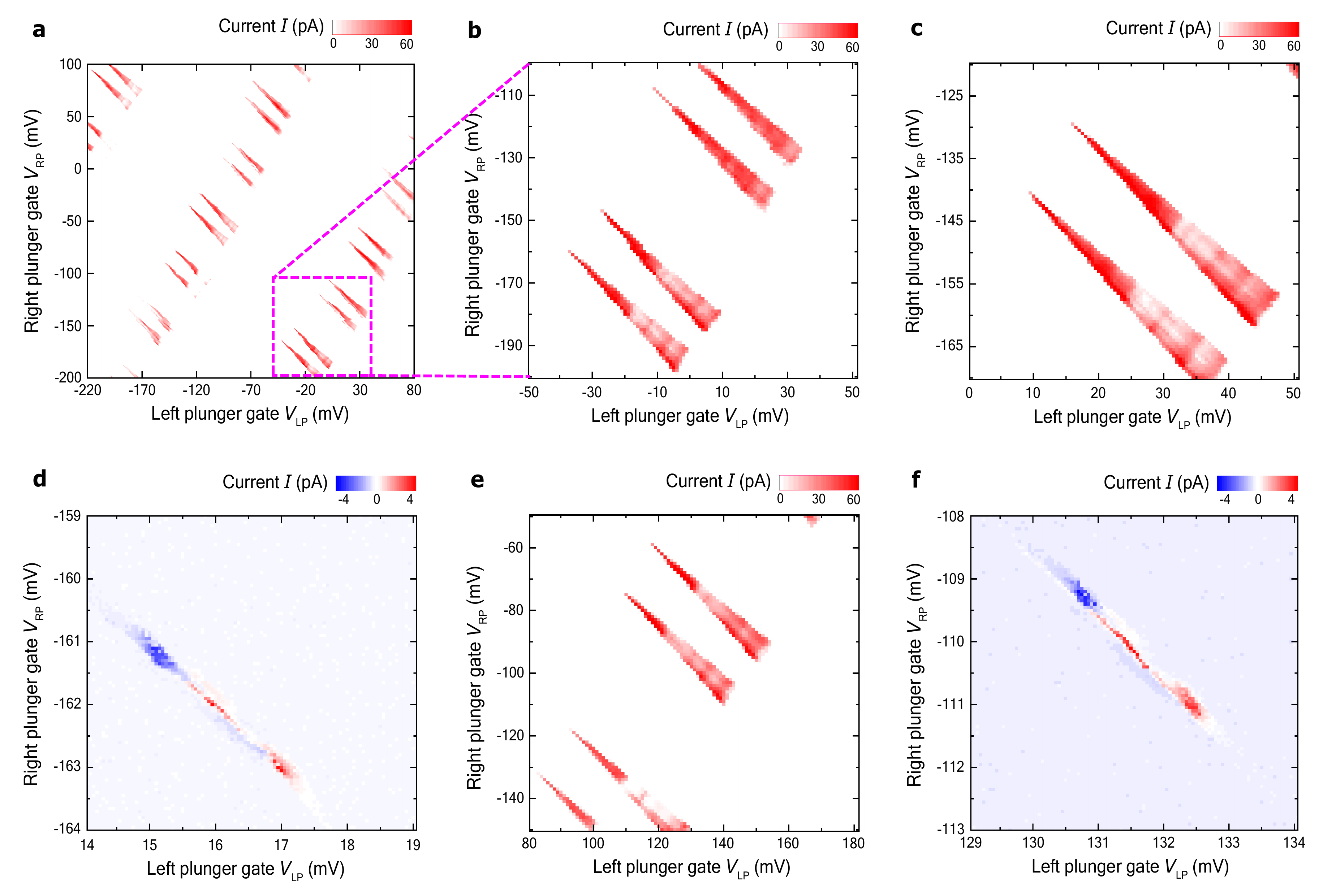}
		\caption{\textbf{a}, Charge stability diagram of the DQD over larger gate voltage range with bias voltage $V_B = \SI{1}{\m V}$. \textbf{b}, Zoom-in of the diagram containing the selected charge configuration at the bottom-most triangle measured at day 35. The corresponding photodetection measurement is presented in Fig. 3a of the main article. \textbf{c}, The same measurement as in b but at day 14 when the configuration was at $V_\mathrm{RP} = \SI{-170}{mV}$ and $V_\mathrm{LP} = \SI{35}{mV}$. \textbf{d}, Photo detection measurement similar to Fig. 3a of the main article done after the measurement in panel c. Panels \textbf{e} and \textbf{f} present a similar set of measurements carried out at day 1 \edit{which is the same day when the phase response measurement of Fig. 3b of the main manuscript was done. These three plots form a set of data where the charge stability diagram (panel \textbf{e} here), photocurrent (panel \textbf{f} here) and phase plot (panel 3b of the main article) are measured with the same gate voltages within a few mV drifts at most between the measurements.}}
		\label{drift}
	\end{center}
\end{figure}

\edit{\subsection*{Charge configuration}}
All the measurements presented in the article and in the Supplemental Material have been performed at the same charge configuration of the DQD. The plunger gate voltages vary from measurement to measurement due to slow drift of the offset charges (in a timescale of a week) that we followed during the course of the measurement. Supplementary Figure~\ref{drift} summarises the measurements that we used for assuring to stay at the same operation point.

Supplementary Figure~\ref{drift}a presents an overview of the different charge numbers. Around each of the configurations, we observe a finite bias triangle pair with unique fingerprint. The magnitude of electrical current varies between each triangle with some having higher current at the base and other at the tip and some on the adjacent sides. Also, each one of them has a characteristic excited state spectrum forming lines along the triangles. The zoom-in of panel b presents the structure of the charge configuration we used throughout the manuscript at around $V_\mathrm{RP} = -\SI{200}{mV}$ and $V_\mathrm{LP} = -\SI{10}{mV}$. Over the course of the measurements, we followed this charge transition point and made control measurement to check that we have the same charge transition whenever the configuration changed considerably. Supplementary Figures~\ref{drift}c and \ref{drift}e show examples of these together with the repetition of the photodetection measurement of Fig. 3a of the main article in panels d and f correspondingly. We see the same features in the finite bias triangles as well as the same photoresponse at $\delta = \pm \delta_r$. Panels d and f have a stronger response to the positive current polarity only around $\delta = 0$ as compared to the data in Fig. 3a of the main article. This arises as in these measurements the offset voltage of the current pre-amplifier was not tuned as well as for the data in the main article. It's worth to note that the magnitude of the photoresponse at $\delta = \pm \delta_r$ is unaltered as it is not sensitive to small bias voltages, i.e. if the chemical potential of either source or drain shift a little in the schematic diagram of Fig. 1a in the main article. \edit{We cannot for sure identify the origin of the current reversing response at $\delta = 0$ of Fig. 3a in the main article. However, we speculate that it arises from a similar effect as reported in Ref.~\citealp{Cornia2019}.}

\begin{figure}[t!]
	\begin{center}
		\centering \includegraphics[width=0.80\textwidth]{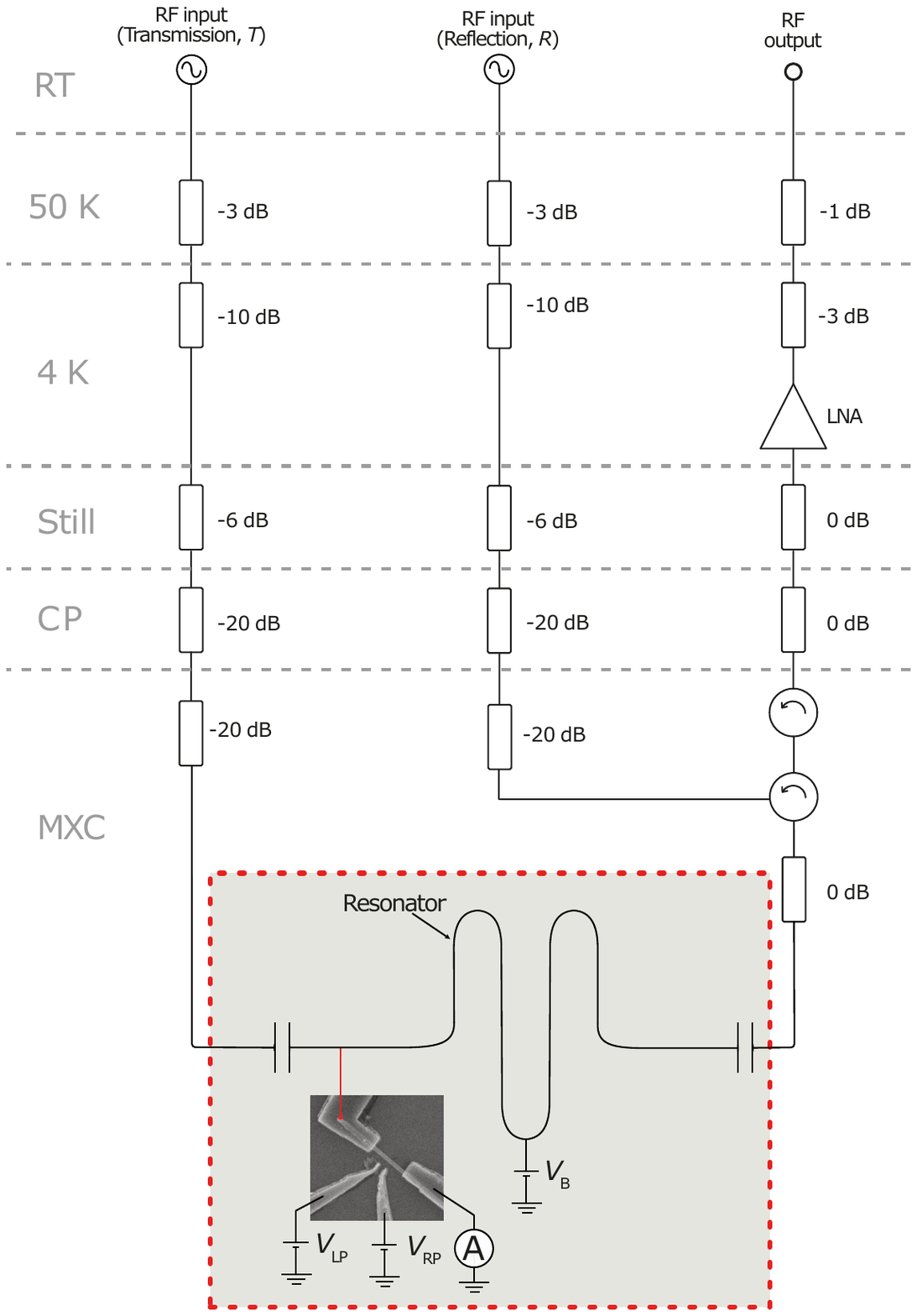}
		\caption {Schematics of RF cables used for sending in microwave photons and measuring the microwave response.}
		\label{rflines}
	\end{center}
\end{figure}

\edit{\subsection*{Determination of the microwave powers at the detector}}

\begin{figure}[t]
	\begin{center}
		\centering \includegraphics[width=0.85\textwidth]{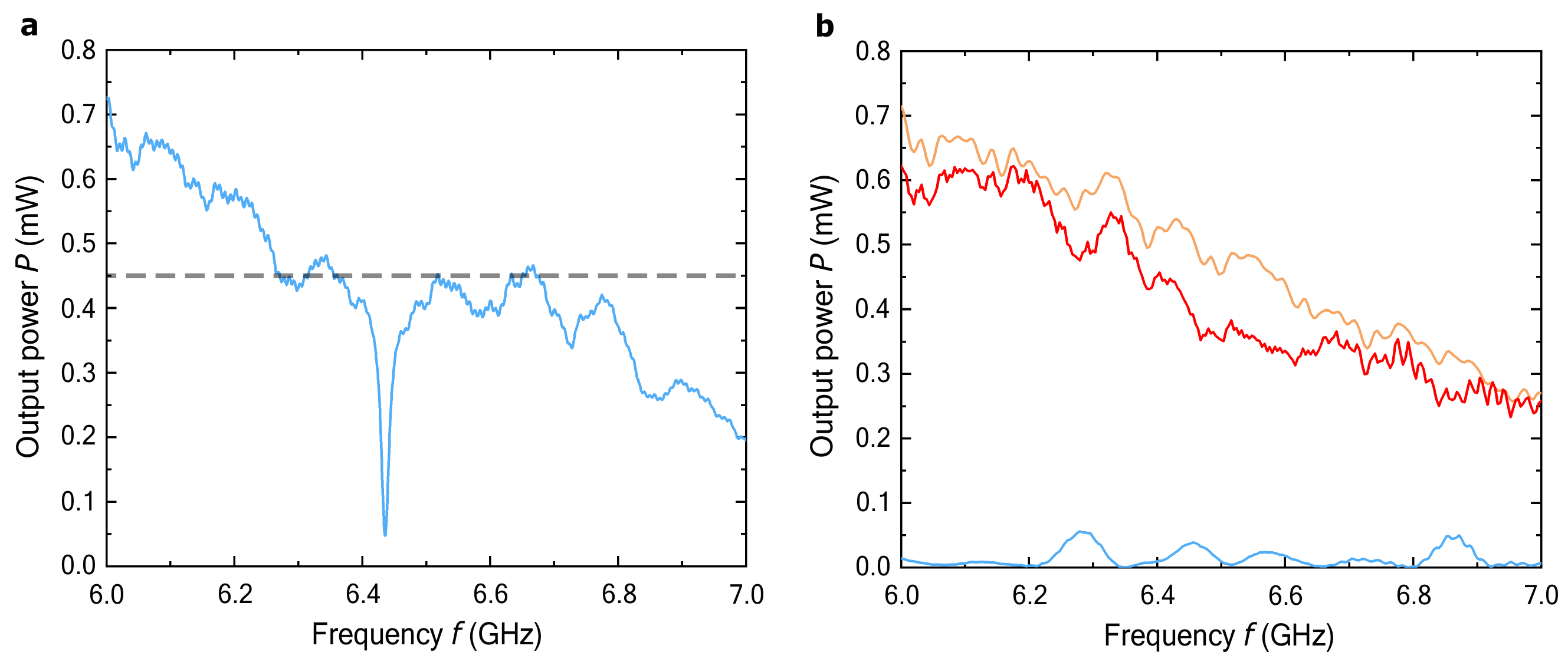}
		\caption{\textbf{a}, Reflected output power $P$ at the output signal digitizer with an input power of 10 mW at the signal generator. In the middle of the figure we see the resonator response and the dashed black line shows the calibration value of -3.5 dBm determined with the background level at $f = 6.3 - 6.6$ GHz. \textbf{b}, Output power $P$ for the reflected (blue) and transmitted signal (red) for a direct transmission line mounted instead of the photodetector. The coral signal shows the reflected power with the sample holding probe completely removed. In that case we obtain somewhat higher signal than the red transmission data as two 30 cm long copper lines and the circuit board holding the device are not part of the signal chain as they are removed with the sample holder.}
		\label{calib}
	\end{center}
\end{figure}

The microwave signals were sent in and the outcoming signals measured with a standard heterodyne detection scheme~\cite{Lang2014}. The cabling in the cryostat is outlined in Supplementary Fig~\ref{rflines}. Outside of the cryostat, the input signal from a RF generator was splitted into two parts. One of the signals was used as a phase reference mixed directly with a local oscillator (LO) and the second one was sent into the device via the cryostat and then mixed with the LO, digitized and then the amplitude and phase response was determined with digital signal processing by determining the amplitudes and phases at the IF frequency.

To determine the in and out going microwave powers at the device, we need to know the microwave attenuation in the signal lines. Supplementary Figure~\ref{rflines} outlines the microwave lines, their attenuators, amplifiers and circulators at each temperature stage. The connecting lines between different temperature stages are nominally identical in all the three lines. In addition to the identical connecting lines, we have the components listed in Supplementary Table~\ref{attenuationTable}. The input lines contain an extra cable, power splitter and attenuators in the cryostat. The output line, on the other hand, has a dual-circulator, cryogenic amplifier, attenuators but with lower losses, two room temperature (RT) amplifiers, mixer, intermediate frequency (IF) attenuation, amplifier and a low pass filter. We list the attenuation/gain for the components in the supplementary table, based on calibration datasheet (cryogenic amplifier) and the datasheet specifications (rest of the components). This leaves the attenuation of the cryostat cables $x$ as the only unknown that is out of construction the same for all lines. 

We determine $x$ close to the measurement frequency by measuring the reflection coefficient well off from the resonance frequency $f_r$ as presented in Supplementary Fig.~\ref{calib}a. The reflection coefficient off from the resonance is unity, i.e. all power is reflected back. By sending in a $P = 10\ \mathrm{dBm}$ microwave signal and measuring a $P = -3.5\ \mathrm{dBm}$ signal at the output yields us a total cable attenuation of $23.9\ \mathrm{dB}$ when accounting for the components of the Supplementary Table ~\ref{attenuationTable}. Since we have identical lines, half of these losses are at the input and half at the output and hence $x = -11.9\ \mathrm{dB}$. The total attenuation from signal generator to the device is hence $80\ \mathrm{dB}$. 

\begin{table}[t]
\centering
\caption{Components and their attenuation or gain in the microwave lines.}
\label{attenuationTable}
\begin{tabular}{ l r }

\hline
 Component & Attenuation/Gain at $f = 7\ \mathrm{GHz}$ (dB) \\ 
\hline 
 {\bf Input lines:} &  \\  
 Output inter-connect cable, Minicircuits 086-36SM+ 	& $-1.84$ \\
 Power splitter, Minicircuits ZX10R-14-S+ & $-7.25$ \\  
 Attenuators in the cryostat (Fig.~\ref{rflines}) & $-59.0$ \\
 Cable losses & $x$ \\ 
 & \\
 {\bf Output line:} &  \\  
 Cryogenic circulator, LNF-CIISC4\_8A, Insertion loss & $-0.2$ \\
 Cryogenic amplifier, LNF-LNC4\_8C & $+39.0$ \\ 
 Cable losses & $x$ \\ 
 Attenuators in the cryostat (Fig.~\ref{rflines}) & $-4.0$ \\
 2 x RT amplifier, Minicircuits ZX60-83LN12+ & $40.4$ \\
 Attenuators at RT & $-6.0$ \\
 Mixer, Minicircuits ZMX-10G+, Conversion loss & $-4.76$ \\
 IF attenuator & $-3.0$ \\
 IF amplifier, Femto DLPVA & $+20.0$ \\
 LP filter attenuation	 & $-3.0$ \\
\hline

\end{tabular}
\end{table}

In addition, to the components listed above, spurious losses could potentially alter the energy balance. The most relevant spurious losses are: 
\begin{itemize}
\item Extra connectors / interconnects in the output line for the circulator and amplifier: We use two Minicircuits 086-4SM+ cables for interconnects, one for the cryogenic amplifier and one for the circulator in the output port. These cables have an insertion loss of $0.2\ \mathrm{dB}$ at $f = 7$ GHz at room temperature. We anticipate that the losses at the operation temperature $T < 3$ K is lower than this as the metals get more conducting towards lower temperatures. Hence these have in total less than $0.4$ dB (i.e. less than $10\ \%$ to the power balance) influence on the power balance calculation above. The extra SMA connectors are specified to have return loss that is better than $30$ dB. Hence the mismatch arising from the extra connectors is negligible (less than $0.1\ \%$ to the power balance).
\item Bonding wires: Bonding wires from the input and output ports to the printed circuit board connect the photodetector to the measurement setup. The bond wires can potentially cause attenuation and impedance mismatch to the circuit. To assess this, we measured a direct transmission line piece bonded instead of the photodetector to the measurement setup. The data is presented in Supplementary Fig.~\ref{calib}b. In this case we anticipate ideally full transmission and no reflected signal. We see that indeed, the transmitted signal in red follows the same background level as the fully reflected background signal in Supplementary Fig.~\ref{calib}a within $20\ \%$ or better. Also, the reflected signal is correspondingly suppressed to  $20\ \%$ or below. Also the reflected signal without the probe (full reflection expected for disconnected probe) presents the same background level with slightly increased amplitude due to two 30 cm cables removed from the signal chain leading to lower overall losses. With these findings, we conclude that bonding wires influence the gain calculation by less than $20\ \%$. It is also worth noting that the two ports are symmetric - also in terms of bonding - and hence the symmetric part of the possible attenuation from the bond wires is accounted for with the above attenuation determination.
\item Impedance mismatches: The input of the cryogenic amplifier has typically some amount of impedance mismatch that gives rise to part of the signal reflected back. Similarly, other components of the setup may cause reflections that give rise to standing waves. The setup is, however, build such that these standing waves are damped. For example for the input of the cryogenic amplifier, the reflected power goes to the circulators. They damp the reflected signal by 20 dB each making the reflected signals negligible. Similarly, the signal reflected from the input to the photodetector is damped by the attenuators of the input line. In the background signals presented in Supplementary Fig.~\ref{calib} we observe that the oscillations arising from standing waves are below $20\ \%$ of the signal power. Similarly, the spuriously reflected signal of Supplementary Fig.~\ref{calib}b is below $20\ \%$ of the input signal. Note that here we have a frequency dependent attenuation that decreases the power towards higher frequencies, which is typical for microwave components. With these arguments, we estimate the total amount of the reflected signals in the measurement chain to be less than $20\ \%$.
\end{itemize}

With the above analysis, the largest uncertainty in the power calibration arises from the $20\ \%$ variations of the background signal level. Therefore we estimate the power calibration to be correct within the $20\ \%$ accuracy, i.e. to about $1\ \mathrm{dB}$. This uncertainty of the input power $P$ appears directly as a uncertainty of the quantum efficiency $\eta$ via Eq.~(1) of the main article. With this, we obtain the error estimate $\eta = 6\ \% \pm 1\ \%$.

In the main article, we have considered the quantum efficiency $\eta$, i.e. the number of electrons passing the device per in-going photons to the input port. Alternatively, it is possible to consider the conversion factor between the measured current $I$ and photon number in the cavity as those have a linear correspondence in the photodetector. Based on the theoretical model, at $P = 1$ fW, we have approximately 15 photons in the cavity with an electrical current of $I = 2.4$ pA. Hence each photon in the cavity contributes to an electrical current of 0.16 pA in the linear response regime.

\bibliographystyle{aipnum4-1}
\bibliography{referencesMicrowavePD}